\documentclass[prb,twocolumn,showpacs,preprintnumbers,amssymb,superscriptaddress]{revtex4}
\usepackage{graphicx}
\usepackage[tbtags]{amsmath}
\usepackage{bm}

\begin{document}

\title
{Low-frequency characterization of quantum tunneling in flux qubits}
\author{Ya.S. Greenberg}
\altaffiliation[On leave from ]{Novosibirsk State Technical
University, 20 K. Marx Ave., 630092 Novosibirsk, Russia.}
\author{A. Izmalkov}
\affiliation{%
Institute for Physical High Technology, P.O. Box 100239, D-07702
Jena, Germany}
\author{M. Grajcar}%
\altaffiliation[On leave from ]{Department of Solid State Physics,
Comenius University, SK-84248 Bratislava, Slovakia.}
\affiliation{Department of Solid State Physics, FSU Jena,
Germany}
\author{E. Il'ichev}
\email{ilichev@ipht-jena.de}
\affiliation{%
Institute for Physical High Technology, P.O. Box 100239, D-07702
Jena, Germany}
\author{ W.~Krech}
\affiliation{Department of Solid State Physics, FSU Jena,
Germany}
\author{H.-G.~Meyer}
\affiliation{%
Institute for Physical High Technology, P.O. Box 100239, D-07702
Jena, Germany}
\author{M.H.S. Amin}%
\affiliation{%
D-Wave Systems Inc., 320-1985 W. Broadway, Vancouver, B.C., V6J
4Y3, Canada}
\author{Alec \surname{Maassen van den Brink}}
\affiliation{%
D-Wave Systems Inc., 320-1985 W. Broadway, Vancouver,
B.C., V6J 4Y3, Canada}

\date{\today}
\begin{abstract} We propose to investigate flux
qubits by the impedance measurement technique (IMT),
currently used to determine the current--phase relation in Josephson
junctions. We analyze in detail the case of a high-quality tank circuit coupled to a persistent-current qubit, to which IMT was
successfully applied in the classical regime. It is shown that low-frequency IMT can give considerable information about the level anticrossing, in particular the value of the tunneling amplitude. An interesting difference exists between applying the ac bias directly to the tank and indirectly via the qubit. In the latter case, a convenient way to find the degeneracy point \emph{in situ} is described. Our design only involves existing technology, and its noise tolerance is quantitatively estimated to be realistic.
\end{abstract}

\maketitle

\section{Introduction}

Josephson-junction flux qubits are known to be candidates for
solid-state quantum computing circuits.\cite{Makhlin} This qubit
variety has good tolerance to external noise, especially to
dangerous background-charge fluctuations.\cite{Blatter} A flux
qubit is a superconducting loop, the two lowest-energy states of
which differ in the direction of circulating persistent current.
For many flux qubits, these two states become degenerate when the
external flux $\Phi_\mathrm{x}$ threading the loop equals
$\Phi_0/2$ ($\Phi_0=h/2e$ is the flux quantum), and quantum
tunneling between them becomes possible. Moving $\Phi_\mathrm{x}$
away from $\Phi_0/2$ lifts the degeneracy and applies a bias
between the two states. When the biasing energy exceeds the
tunneling amplitude~$\Delta$ the tunneling stops, but the relative
phase between the two states will still evolve in time. This,
together with coherent tunneling, provides single-bit quantum gate
operations. To have a universal set of gates, necessary for
quantum computing, one needs to be able to couple two qubits. The
methods of coupling two flux qubits and performing gate operations
are beyond the present scope. Instead, we propose a method to
characterize the quantum behavior of a flux qubit by coupling it
to a tank circuit. The discussion will be quite general and can be
applied to different types of flux qubit such as
rf-SQUID,\cite{rfsquid} three-Josephson-junction
(3JJ),\cite{Mooij,Orlando} multi-terminal,\cite{amin02} etc. We
will use the example of the 3JJ qubit, where quantum superposition
of the macroscopic current states has been observed.\cite{Mooij}

\begin{figure}
\includegraphics[width=7cm]{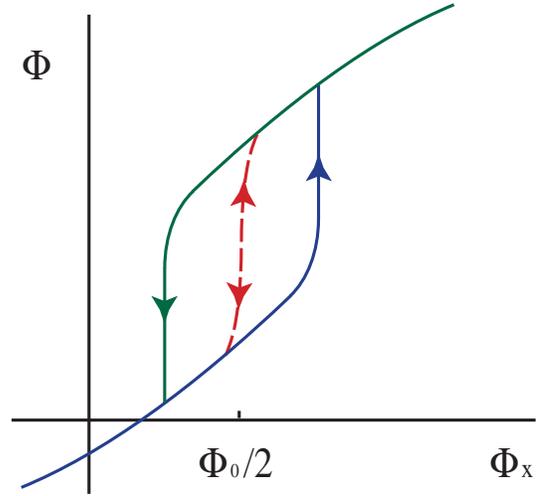}
\caption{\label{fig1} Solid lines: hysteretic dependence of the
total flux~$\Phi$ on the external flux $\Phi_\mathrm{x}$ in the
classical regime. Dashed line: disappearance of the hysteresis by
quantum tunneling.}
\end{figure}

Due to the loop self-inductance, the total qubit flux~$\Phi$ may
differ from $\Phi_\mathrm{x}$, depending on the direction of the
persistent current. Figure~\ref{fig1} shows the
$\Phi$--$\Phi_\mathrm{x}$ curve for a typical flux qubit. The
solid lines correspond to classical behavior. Near the degeneracy
point, the diagram is hysteretic, a signature of the qubit's
bistability. This has been observed for the 3JJ
in Refs.~\onlinecite{Tanaka,APL7}. In the quantum regime,
tunneling between the states at degeneracy may eliminate the
hysteresis (dashed line in Fig.~\ref{fig1}). This phenomenon will
be discussed in detail below.

\begin{figure}[ht]
\includegraphics[width=8cm]{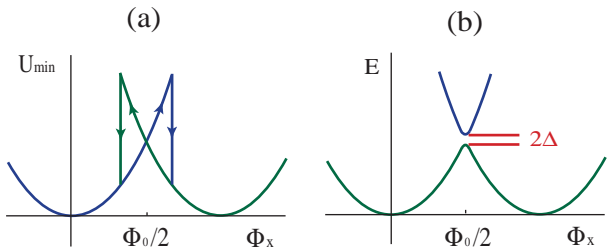} \caption[]{\small
(a) Minimum energies of a qubit as a function of external magnetic
flux in the classical regime. (b) Quantum mechanical energy profile
for the same qubit as in (a).} \label{fig2}
\end{figure}

In general, one can plot the classical (local) minimum
energies of a flux qubit as in Fig.~\ref{fig2}a. The left (right)
branch then corresponds to (counter\nobreakdash-)clockwise
flow of the spontaneous current. The
hysteresis is also evident from this diagram. In the
quantum regime, there will be discrete local states in each of
the qubit's bistable potential wells. From now on we denote the
lowest-lying such states as $\Psi^\mathrm{l}$ and $\Psi^\mathrm{r}$, corresponding
to ``left" and ``right" directions of the persistent current
respectively. At $\Phi_\mathrm{x}=\Phi_0/2$, resonant
tunneling will render the lowest eigenstates of the full
Hamiltonian as superpositions $(\Psi^\mathrm{l}{\pm}\Psi^\mathrm{r})/\sqrt{2}$. A
small splitting equal to $2\Delta$ will appear between their
energies (Fig.~\ref{fig2}b). Starting with the qubit in its
ground state (lower band in Fig.~\ref{fig2}b), adiabatically
changing $\Phi_\mathrm{x}$ will keep it in the ground state. This
means that by passing through the degeneracy point, the qubit
will continuously transform from $\Psi^\mathrm{l}$ to $\Psi^\mathrm{r}$. This
pure quantum behavior is shown by the dashed line in
Fig.~\ref{fig1}. On the other hand, if $\Phi_\mathrm{x}$ changes
rapidly, there is a considerable probability to excite the qubit
and therefore continue on the same classical branch (left or
right). This so-called Landau--Zener effect can be used to
distinguish the classical from the quantum energy curves.

\begin{figure}[h]
\includegraphics[width=8cm]{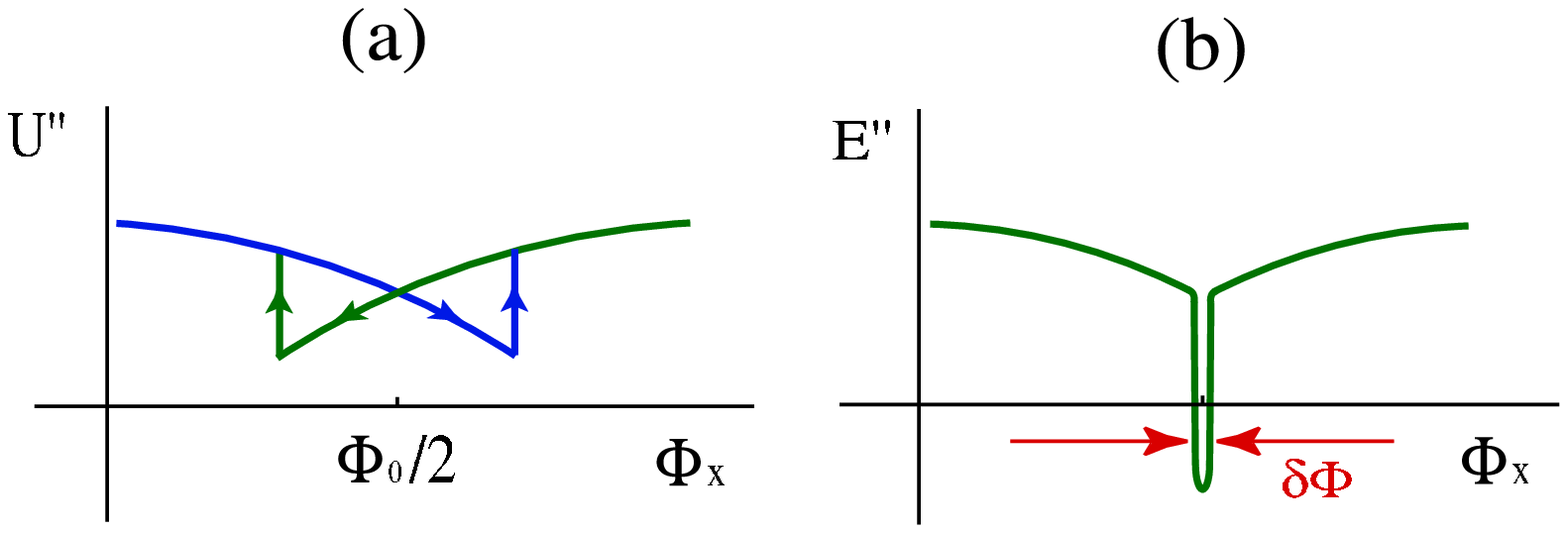} \caption{\small
(a) Second derivative of the qubit's classical minimum energy
vs external magnetic flux. (b)~Second derivative of the same
qubit's ground-state energy in the quantum regime.} \label{fig3}
\end{figure}

The curvature of the energy profile is related to the qubit's
effective inductance and is therefore important for measurement.
Figure~\ref{fig3} displays the second derivative of the curves in
Fig.~\ref{fig2}. In the classical regime (Fig.~\ref{fig3}a), the
hysteretic behavior is the same as for the energy. On the other
hand, in the quantum regime the hysteresis is replaced by a sharp
spike, due to the level anticrossing. The appearance of this
spike can be ascribed to enhanced susceptibility of the system
due to tunneling. Its size and width can provide information
about $\Delta$.

\begin{figure}
\includegraphics[width=8cm]{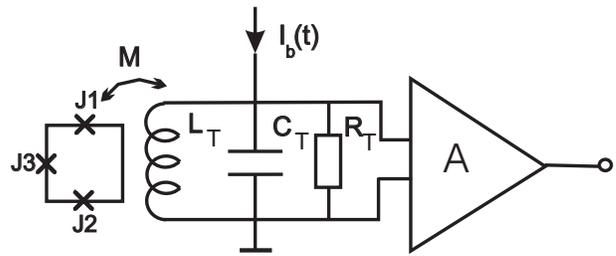}
\caption{\small Flux qubit coupled to a tank: direct
biasing scheme.}\label{fig4}
\end{figure}

A simple experimental implementation is to inductively couple the
qubit to an $LC$ tank circuit with known inductance
$L_\mathrm{T}$, capacitance $C_\mathrm{T}$, and quality $Q$
through a mutual inductance $M$ (Fig.~\ref{fig4}). The resonant
characteristics of the tank circuit (frequency, phase shift, etc.)
will then be sensitive to the qubit inductance and therefore to
its energy curvature. In particular, the spike in Fig.~\ref{fig3}b
appears as sharp dips in both phase shift and tank voltage as a
function of $\Phi_\mathrm{x}$ (see Section~\ref{bit-tank}).

This method, known as \textit{impedance measurement technique}
(IMT), has been used for current--phase measurements of Josephson
junctions. It originates from the pioneering work of Rifkin and
Deaver,\cite{Rifkin-Deaver} and is analyzed in detail in
Ref.~\onlinecite{IMT-2}. IMT has also successfully been applied to
a 3JJ qubit in the classical regime,\cite{APL7} and the hysteretic
dependence of the ground-state energy on $\Phi_\mathrm{x}$ (cf.\
Fig.~\ref{fig2}a) was observed as predicted in
Ref.~\onlinecite{Orlando}. The method has also been used for the
investigation of quantum transitions in an rf-SQUID
(Ref.~\onlinecite{Clark} and references therein).

First of all, in Section~\ref{3dd-dyn} we calculate the two qubit energies in more detail than in Ref.~\onlinecite{Orlando}. In
Section~\ref{bit-tank}, we study the qubit's interaction with a
high-quality resonant tank, showing that low-frequency IMT yields useful information about
the qubit's quantum behavior. Finally, in
Section~\ref{noise-src}, the effect of noise is considered.

\section{Quantum dynamics of the 3JJ qubit}
\label{3dd-dyn}

The 3JJ qubit\cite{Orlando} consists of three Josephson junctions
in a loop with very small inductance~$L$, typically in the pH
range. This insures effective decoupling from the environment. Two
junctions have equal critical current $I_\mathrm{c}$ and
(effective) capacitance~$C$, while those of the third junction are
slightly smaller: $\alpha I_\mathrm{c}$ and $\alpha C$, with
$0.5<\alpha<1$. If the Josephson energy
$E_\mathrm{J}=I_\mathrm{c}\Phi_0/2\pi$ is much larger than the
Coulomb energy $E_C=e^2\!/2C$, the Josephson phase is well
defined. Near $\Phi_\mathrm{x}=\Phi_0/2$, this system has two
low-lying quantum states.\cite{Orlando, Mooij1} The energy
splitting between them in the presence of a small flux bias has
been given in Ref.~\onlinecite{Mooij}, but only for a particular
choice of, e.g., $\alpha$ and $g\equiv E_\mathrm{J}/E_C$. In this
section we derive the splitting with its explicit dependence on
the qubit parameters. The energy levels are derived from the
Hamiltonian [see Eq.~(12) in Ref.~\onlinecite{Orlando}]
\begin{equation}
  H_0 = \frac{P_\varphi^2}{2M_\varphi}+\frac{P_\theta^2}{2M_\theta}
  + U(f_\mathrm{x},\varphi,\theta)\;,\label{defH0}
\end{equation}
where $\varphi = (\varphi_1 {+} \varphi_2)/2$, $\theta = (\varphi_1
{-} \varphi_2)/2$ with $\varphi_{1,2}$ the phase
differences across the two identical junctions,
$P_\varphi=-i\hbar\partial_\varphi$,
$P_\theta=-i\hbar\partial_\theta$,
$M_\varphi=(\Phi_0/2\pi)^22C$,
$M_\theta=(1{+}2\alpha)M_\varphi$, and
\begin{equation}
  U(f_\mathrm{x},\varphi,\theta) = E_\mathrm{J}
  [\alpha-2\cos\varphi\cos\theta +\alpha \cos ( 2\pi\!f_\mathrm{x}{+}2\theta) ]\;.\label{defU}
\end{equation}
In contrast to Ref.~\onlinecite{Orlando}, we define the flux bias
$f_\mathrm{x}=\linebreak\Phi_\mathrm{x}/\Phi_0-\frac{1}{2}$ as a small deviation from degeneracy.

Since the qubit is assumed to have small $L$ and $I_\mathrm{c}$
(typically $L\approx10$~pH, $I_\mathrm{c}\approx100$~nA), the
shielding factor $LI_\mathrm{c}/\Phi_0\approx 0.001$. Hence, in
(\ref{defH0}) we have neglected the shielding current, considering
$\Phi$ as an external flux.

At $f_\mathrm{x}=0$, the potential~(\ref{defU}) has two minima at
$\varphi=0$, $\theta=\pm\theta_*$, with $\cos\theta_*=1/2\alpha$
($\theta_*>0$). Tunneling lifts their degeneracy, leading to
energy levels $E_\pm=\varepsilon_0 \pm\nobreak \Delta$.  To find
the levels for $|f_\mathrm{x}|\ll1$ we expand Eq.~(\ref{defU})
near its minima, retaining linear terms in $f_\mathrm{x}$ and
quadratic terms in $\phi,\theta$. Define $\theta^\mathrm{r/l}_*$
as the minima, shifted due to $f_\mathrm{x}$:
\begin{equation}
  \theta^\mathrm{r/l}_*=\pm\theta_*+2\pi\!f_\mathrm{x}\frac{1{-}2\alpha^2}{4\alpha^2{-}1}\;;
\end{equation}
that is, the upper (lower) sign refers to the right
(left) well. The potential energy then reads
\begin{align}
  \frac{U}{E_\mathrm{J}}&= -\frac{1}{2\alpha}\mp
  f_\mathrm{x}\frac{\pi}{\alpha}\sqrt{4\alpha^2{-}1}
  +\frac{\varphi^2}{2\alpha}\left(1\pm2\pi\!
  f_\mathrm{x}\frac{2\alpha^2-1}{\sqrt{4\alpha^2{-}1}}\right)\notag\\
  &\quad +{(\theta-\theta^\mathrm{r/l}_*)}^2\left(2\alpha-\frac{1}{2\alpha}\pm
  f_\mathrm{x}\frac{\pi}{\alpha}\frac{2\alpha^2{+}1}{\sqrt{4\alpha^2{-}1}}\right).
\end{align}

Near degeneracy, the eigensolutions of $H_0\Psi_\pm=E_\pm\Psi_\pm$
can be written as superpositions
$\Psi_\pm=a_\pm\Psi^\mathrm{l}+b_\pm\Psi^\mathrm{r}$, yielding the
well-known eigenenergies
$E_\pm=(\varepsilon^\mathrm{l}{+}\varepsilon^\mathrm{r})/2\pm
\sqrt{(\varepsilon^\mathrm{l}{-}\varepsilon^\mathrm{r})^2\!/4+\Delta^2}$,
with
$\varepsilon^\mathrm{r/l}=\langle\Psi^\mathrm{r/l}|H_0|\Psi^\mathrm{r/l}\rangle$.
The matrix element $\Delta$ cannot accurately be found in terms
of~$\Psi^\mathrm{r/l}$. In what follows it is assumed constant,
\begin{multline}
  \Delta=2E_\mathrm{J}\sqrt{\frac{2\alpha{-}1}{\alpha g}}\\
  \times\exp\!\left[\sqrt{\frac{g(2\alpha{+}1)}{\alpha}}\left(\arccos\frac{1}{2\alpha}
  -\sqrt{4\alpha^2{-}1}\right)\!\right],
\end{multline}
neglecting its dependence on~$f_\mathrm{x}$.

To find the dependence of $E_\pm$ on $f_\mathrm{x}$, we take
$\Psi^\mathrm{r/l}$ to be oscillator ground states in their
respective wells:
\begin{equation}\begin{split}
  \Psi^\mathrm{r/l}&=
  \frac{1}{\sqrt{\hbar\pi}}\left(M_\varphi\omega_\varphi^\mathrm{r/l}
  M_\theta\omega_\theta^\mathrm{r/l}\right)^{\!1/4} \\
  &\quad\times\exp\biggl(-\frac{M_\varphi\omega_\varphi^\mathrm{r/l}}{2\hbar}\varphi^2
  -\frac{M_\theta\omega_\theta^\mathrm{r/l}}{2\hbar}
  {(\theta{-}\theta_*^\mathrm{r/l})}^2\biggr)\;,
\end{split}\end{equation}
corresponding to
\begin{equation}
  \varepsilon^\mathrm{r/l}=E_\mathrm{J}\left(-\frac{1}{2\alpha}\mp
  f_\mathrm{x}\frac{\pi}{\alpha}\sqrt{4\alpha^2{-}1}\right)
  +\frac{\hbar\omega_\varphi^\mathrm{r/l}}{2}+\frac{\hbar\omega_\theta^\mathrm{r/l}}{2}\;,
\end{equation}
where
\begin{align}
  \hbar\omega_\varphi^\mathrm{r/l} &= E_\mathrm{J}\sqrt{\frac{4}{\alpha
  g}}\left(1\pm\pi f_\mathrm{x}\frac{2\alpha^2{-}1}{\sqrt{4\alpha^2{-}1}}\right),\\
  \hbar\omega_\theta^\mathrm{r/l} &=
  E_\mathrm{J}\sqrt{\frac{4(2\alpha{-}1)}{\alpha g}}\left(1\pm\pi
  f_\mathrm{x}\frac{2\alpha^2{+}1}{\left(4\alpha^2{-}1\right)^{3/2}}\right).
\end{align}

Combining the above, one finds the eigenenergies
\begin{equation}\label{E_pm}
  E_\pm=\varepsilon_0\pm\sqrt{E_\mathrm{J}^2f_\mathrm{x}^2\lambda^2(\alpha)+\Delta^2}\;,
\end{equation}
where
\begin{gather}
  \varepsilon_0=E_\mathrm{J}\left(-\frac{1}{2\alpha}
  +\frac{1+\sqrt{2\alpha{-}1}}{\sqrt{\alpha g}}\right),\label{e_o}\\
  \begin{split}
  \frac{\alpha}{\pi}\lambda(\alpha)&=
  \sqrt{\frac{\alpha}{g}}\left(\frac{2\alpha^2{-}1}{\sqrt{4\alpha^2{-}1}}
  +\frac{2\alpha^2{+}1}{\sqrt{2\alpha{+}1}(4\alpha^2{-}1)}\right)\\
  &\quad-\sqrt{4\alpha^2{-}1}\;.\label{lambda}\end{split}
\end{gather}
The splitting given by Eq.~(\ref{E_pm}) differs from
that of Eq.~(1) in Ref.~\onlinecite{Mooij} by a factor
$\lambda(\alpha)$ which explicitly accounts for the dependence of
$E_\pm$ on $\alpha$ and~$g$.

For stationary states, the current in the qubit loop
can be calculated either as the average of the current operator
$\hat{I}_\mathrm{q}=I_\mathrm{c}\sin(\varphi+\theta)$ over
the eigenfunctions, or as the derivative of the energy over
the external flux:
\begin{equation}\label{I_q}
  I_\mathrm{q}=\langle\Psi_\pm|\hat{I}_\mathrm{q}|\Psi_\pm\rangle=\frac{\partial
  E_\pm}{\partial\Phi}=\pm
  I_\mathrm{c}f_\mathrm{x}\frac{\lambda^2(\alpha)}{\pi}\frac{E_\mathrm{J}}{\hbar\omega_0}\;,
\end{equation}
where $\hbar\omega_0=E_+-E_-$. In equilibrium at finite temperature~$T$, Eq.~(\ref{I_q})
readily generalizes to
\begin{equation}\begin{split}
  I_\mathrm{q}&=\langle\Psi_+|\hat{I}_\mathrm{q}|\Psi_+\rangle\rho_{++}^\mathrm{eq}
  +\langle\Psi_-|\hat{I}_\mathrm{q}|\Psi_-\rangle\rho_{--}^\mathrm{eq}\\
  &=-I_\mathrm{c}\frac{E_\mathrm{J}f_\mathrm{x}\lambda^2(\alpha)}{\pi\hbar\omega_0}
  \tanh\!\left(\frac{\hbar\omega_0}{2k_\mathrm{B}T}\right)\label{I_qrho}
\end{split}\end{equation}
with the density matrix elements
$\rho_{++}^\mathrm{eq}=e^{-E_+/k_\mathrm{B}T}\!/Z$ and
$\rho_{--}^\mathrm{eq}=e^{-E_-/k_\mathrm{B}T}\!/Z$, where
$Z=e^{-E_+/k_\mathrm{B}T}+e^{-E_-/k_\mathrm{B}T}$.

\section{Qubit--tank interaction}
\label{bit-tank}

We propose here to extract information about the quantum dynamics
of a flux qubit with the aid of a classical linear high-quality
tank circuit, coupled to the qubit via a mutual inductance~$M$.
The tank consists of a capacitor $C_\mathrm{T}$, inductor
$L_\mathrm{T}$, and a resistor $R_\mathrm{T}$ which are connected
in parallel and driven by a current source $I_\mathrm{b}(t)$
(Fig.~\ref{fig4}). The problem of coupling a quantum object to a
dissipative classical one has no unique theoretical solution.
However, if we assume that the classical object is much slower
than the quantum one we may solve for the latter's motion,
accounting for the coupling coordinates of the former as mere
external parameters.\cite{Clark} Here, the characteristic
frequency $\Delta/h$ of the qubit is in the GHz range, while the
resonances $\omega_\mathrm{T}$ of our tank circuit lie below 100
MHz. There exist two different schemes of coupling a tank
circuit to the qubit. First we consider direct biasing, where a
current $I_\mathrm{b}(t)=I_0\cos\omega t$ is fed directly into
$L_\mathrm{T}$ (Fig.~\ref{fig4}).

\subsection{Direct biasing scheme}

The voltage across the tank circuit evolves as
\begin{equation}\label{q-tank}
  \ddot{V}+\frac{\omega_\mathrm{T}}{Q}\dot{V}+\omega_\mathrm{T}^2 V =
  -M\omega_\mathrm{T}^2\dot{I}_\mathrm{q}+\frac{1}{C_\mathrm{T}}\dot{I}_b(t)\;.
\end{equation}
Here, $Q=\omega_\mathrm{T}R_\mathrm{T}C_\mathrm{T} \gg 1$ and
$\omega_\mathrm{T}=1/\sqrt{L_\mathrm{T}C_\mathrm{T}}$;
$I_\mathrm{q}$ is given by (\ref{I_q}) or (\ref{I_qrho}), and
depends on the qubit flux $\Phi=\Phi_\mathrm{x}+MI_L$, where
$I_L=L_\mathrm{T}^{-1}\int V\,dt$ is the current in $L_\mathrm{T}$
and $\Phi_\mathrm{x}$ is time independent. Below we study the
simplest case $k_\mathrm{B}T\ll\Delta$, so that the qubit is
definitely in its ground state $E_-$. Then, Eq.~(\ref{q-tank})
takes the form
\begin{equation}\label{V-dyn}
  \ddot{V}+\frac{\omega_\mathrm{T}}{Q}\dot{V}+\omega_\mathrm{T}^2 V =
  -k^2L\omega_\mathrm{T}^2\frac{d^2E_-}{d\Phi^2}V+\frac{1}{C_\mathrm{T}}\dot{I}_b(t)
\end{equation}
where $k^2\equiv M^2\!/LL_\mathrm{T}$,
\begin{equation}
  \frac{d^2E_-}{d\Phi^2}=\frac{I_\mathrm{c}^2\Delta^2\lambda^2(\alpha)}
  {(2\pi)^2\left(E_\mathrm{J}^2f^2\lambda^2(\alpha)+\Delta^2\right)^{3/2}}\;,\label{d2E}
\end{equation}
and $f=[\Phi_\mathrm{x}{+}MI_L(t)]/\Phi_0 - \frac{1}{2}$ ($|f|\ll1$). Thus,
(\ref{V-dyn}) is nonlinear in $V$. Since the coupling
to the qubit is small one may apply the method of harmonic balance,
which is well known in rf-SQUID theory.\cite{Likharev}
Accordingly, if $\omega\approx\omega_\mathrm{T}$, then
$V$ oscillates with frequency~$\omega$, while its amplitude $v$ and phase
$\chi$ are slow functions of time: $V(t)=v(t)\cos[\omega
t+\chi(t)]$. From (\ref{V-dyn}) we obtain
\begin{gather}
  \dot{v}=-\frac{\omega_\mathrm{T}v}{2Q}+\frac{I_0\cos\chi}{2C_\mathrm{T}}\label{a-dot}\\
  \dot{\chi}=\omega_\mathrm{T}\xi_0-\frac{I_0\sin\chi}{2vC_\mathrm{T}}
  -\frac{k^2\omega_\mathrm{T}LI_\mathrm{c}^2}{2\Delta}
  \left(\frac{\lambda(\alpha)}{2\pi}\right)^{\!2}\!F(v,f_\mathrm{x})\label{th-dot}
\end{gather}
with the detuning $\xi_0=(\omega_\mathrm{T}{-}\omega)/\omega_\mathrm{T}$, and where
\begin{equation}
  F(v,f_\mathrm{x})=\frac{1}{\pi}\int_0^{2\pi}\!\!d\phi\,
  \frac{\cos^2\phi}{\bigl[1+\eta^2\left(f_\mathrm{x}+\gamma\sin\phi\right)^2\bigr]^{3/2}}\;,
\end{equation}
with $\eta=E_\mathrm{J}\lambda(\alpha)/\Delta$ and
$\gamma=Mv/\omega_\mathrm{T}L_\mathrm{T}\Phi_0$.

\begin{figure}
\includegraphics[width=7cm]{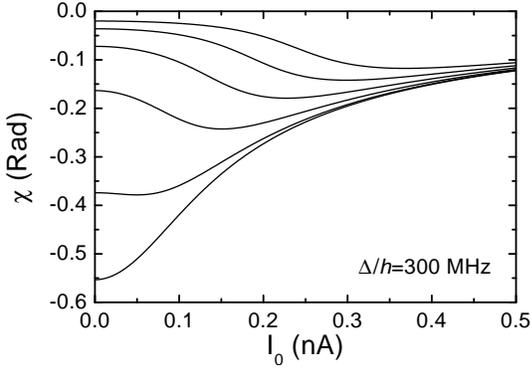}
\caption{\label{fig5} Tank phase $\chi$ vs bias
amplitude $I_0$; $\Delta/h=300$~MHz. From the lower to the upper curve, the bias flux
$10^4\!f_\mathrm{x}$ takes the values $0, 2, 4, 6, 8, 10$.}
\end{figure}

\begin{figure}
\includegraphics[width=7cm]{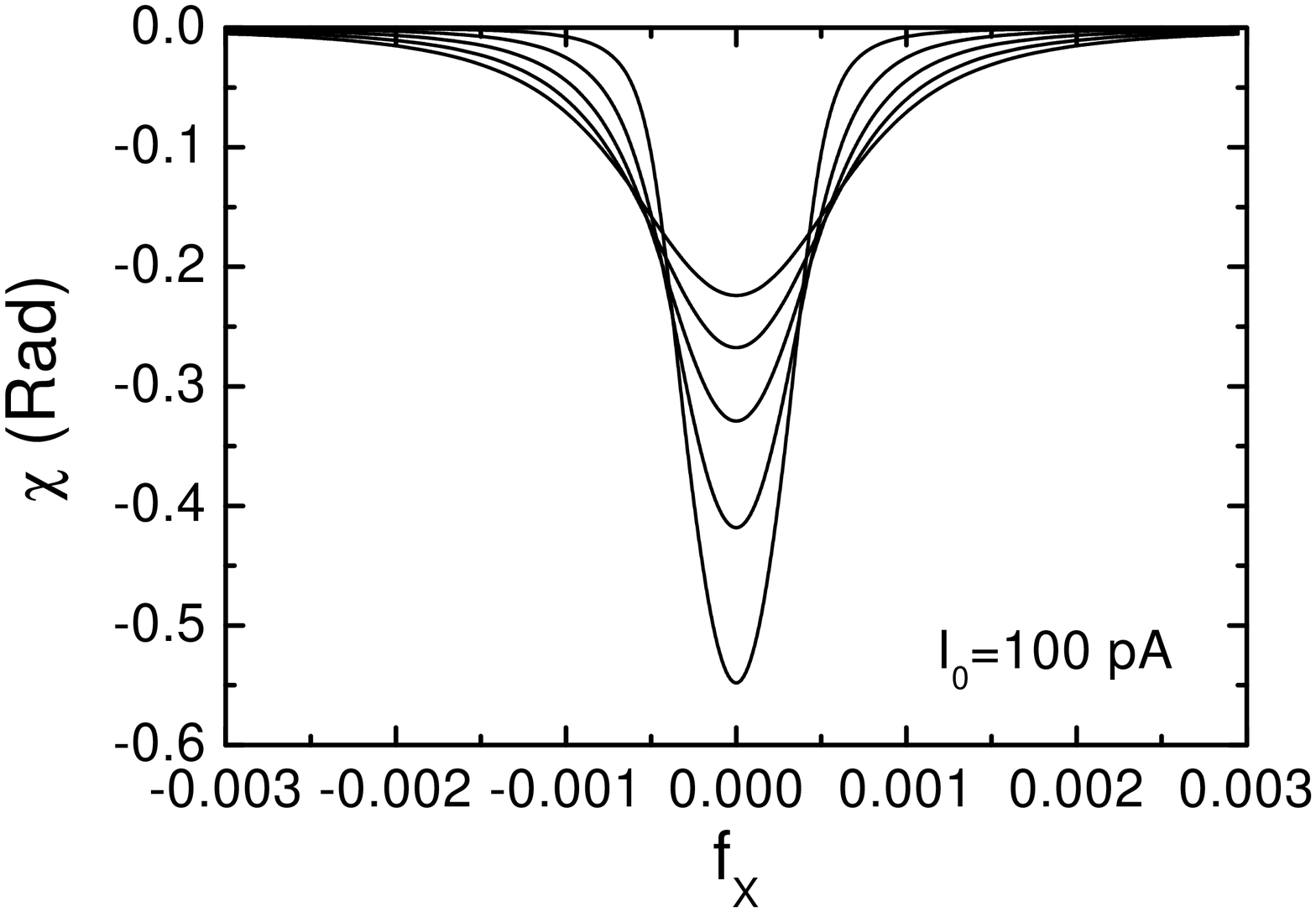}
\caption{\label{fig6} Tank phase $\chi$ vs bias flux
$f_\mathrm{x}$; $I_0=100$ pA. From the lower to the upper curve
(at $f_\mathrm{x}=0$), the tunneling frequency $\Delta/h$ takes
the values $150, 300, 450, 600, 750$~MHz.}
\end{figure}

Setting $\dot{v}=\dot{\chi}=0$ in (\ref{a-dot}) and (\ref{th-dot}) one obtains
the stationary tank voltage and phase,
\begin{gather}
  v^2\left(1+4Q^2\xi^2(v,f_\mathrm{x})\right)
    =I_0^2\omega_\mathrm{T}^2L_\mathrm{T}^2Q^2\label{V-I}\\
  \tan\chi=2Q\xi(v,f_\mathrm{x})\;,\label{phi-I}
\end{gather}
where we introduced a flux-dependent detuning
\begin{equation}
  \xi(v,f_\mathrm{x})=\xi_0-k^2\frac{LI_\mathrm{c}^2}{2\Delta}
  \left(\frac{\lambda(\alpha)}{2\pi}\right)^{\!2}F(v,f_\mathrm{x})\;.\label{xi}
\end{equation}

\begin{figure}
\includegraphics[width=7cm]{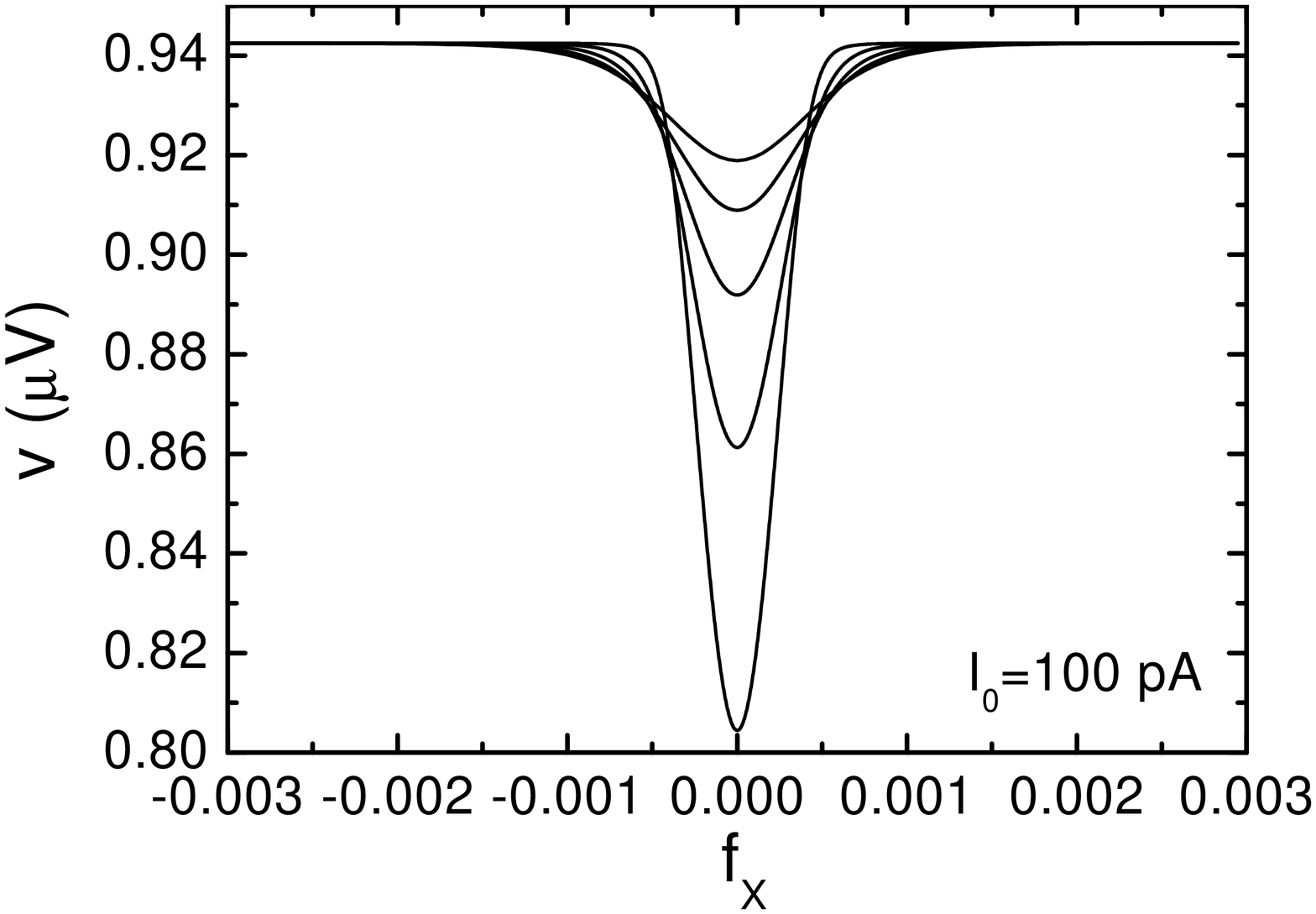}
\caption{\label{fig7} Tank voltage $v$ vs bias flux $f_\mathrm{x}$; $I_0=100$ pA.
From the lower to the upper curve (at $f_\mathrm{x}=0$),
the tunneling frequency $\Delta/h$ takes the values $150, 300, 450, 600, 750$~MHz.}
\end{figure}

\begin{figure}[t]
\includegraphics[width=7cm]{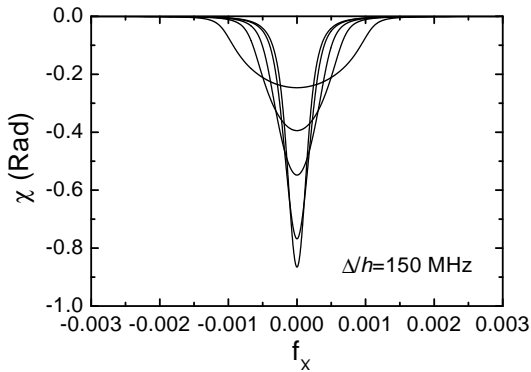}
\caption{\label{fig8} Tank phase $\chi$ vs
bias flux $f_\mathrm{x}$; $\Delta/h=150$~MHz. From the lower to the upper curve
(at $f_\mathrm{x}=0$), the bias amplitude $I_0$ takes the values $10, 50,
100, 150, 250$ pA.}
\end{figure}

\begin{figure}
\includegraphics[width=7cm]{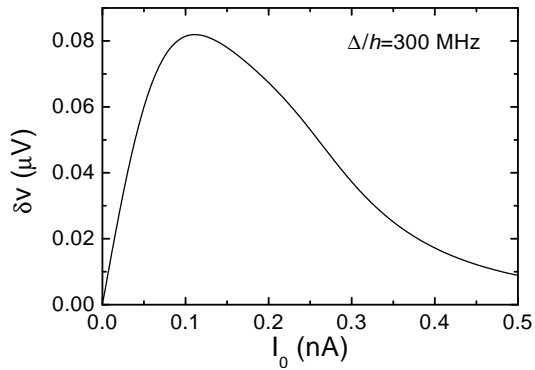}
\caption{\label{fig9}  Voltage modulation
$\delta v=v(f_\mathrm{x}{=}0)-v(f_\mathrm{x}{=}10^{-3})$ vs bias current
$I_0$; $\Delta/h=300$~MHz.}
\end{figure}

We have used Eqs.\ (\ref{V-I})--(\ref{xi}) to
find voltage--flux $v(f_x)$, phase--current $\chi(I_0)$, and
phase--flux $\chi(f_\mathrm{x})$ characteristics at resonance
$\omega=\omega_\mathrm{T}$. We take the qubit parameters
$I_\mathrm{c}=400$ nA, $\alpha=0.8$, $L=15$ pH, $g=100$, a tank
with $L_\mathrm{T}=50$~nH, $Q=1000$, $\omega_\mathrm{T}/2\pi=30$
MHz, and $k=10^{-2}$. The $\chi(I_0)$ curves for several
$f_\mathrm{x}$ are shown in Fig.~\ref{fig5}. The
$\chi(f_\mathrm{x})$ and $v(f_x)$ curves are shown in Figs.\
\ref{fig6} and~\ref{fig7} for various $\Delta$. The sharp dips in
Figs.\ \ref{fig6} and~\ref{fig7} correspond to the spike in the second
derivative of the energy profile in Fig.~\ref{fig3}b. Clearly, the
width of the dips is correlated with $\Delta$: with the increase
of $\Delta$ the width of the dips also increases. The
$\chi(f_\mathrm{x})$ curves for different $I_0$ are shown in
Fig.~\ref{fig8}. The shape and the value of $\chi$ are seen to be
very sensitive to $I_0$. The dependence of the voltage modulation
$\delta v\equiv v(f_\mathrm{x}{=}0)-v(f_\mathrm{x}{=}10^{-3})$ on
$I_0$ is shown in Fig.~\ref{fig9}.

\subsection{Scheme with separate driving coil}

\begin{figure}[t]
\includegraphics[width=7cm]{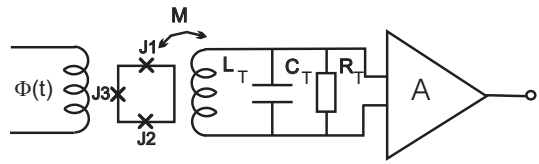}
\caption{\label{fig10} Flux qubit coupled to a tank:
scheme with a separate driving coil.}
\end{figure}

In this scheme, a bias flux $\Phi_\mathrm{b}(t)=\Phi_\mathrm{ac}\sin\omega t$
is applied to the qubit loop from a separate coil (Fig.~\ref{fig10}).
The tank response is similar to (\ref{V-dyn}):
\begin{equation}
 \ddot{V}+\frac{\omega_\mathrm{T}}{Q}\dot{V}+\omega_\mathrm{T}^2 V =\left[-
  M\omega_\mathrm{T}^2\frac{d^2E_-}{d\Phi^2}\Phi_\mathrm{ac}
  +\widetilde{\Phi}_\mathrm{ac}\right]\omega\cos\omega t\;,\label{V-dyn2}
\end{equation}
where $\widetilde{\Phi}_\mathrm{ac}$ is the flux which the external coil
couples directly into the tank and $d_\Phi^2E_-$ is given by
(\ref{d2E}) with $f=(\Phi_\mathrm{x}{+}\Phi_\mathrm{ac}\sin\omega t)/\Phi_0 -
\frac{1}{2}\equiv f_\mathrm{x}+f_\mathrm{ac}\sin\omega t$. Rewriting the
first term on the rhs of (\ref{V-dyn2}) as
\begin{equation}
  -k\frac{I_\mathrm{c}^2}{\Delta}\sqrt{L_\mathrm{T}L}\left(\frac{\lambda}{2\pi}\right)^{\!2}
  \omega_\mathrm{T}^2\omega\Phi_0f_\mathrm{ac}G(t)
\end{equation}
makes its time dependence manifest:
\begin{equation}
  G(t)=\frac{\cos\omega t}{\bigl[1+\eta^2\left(f_\mathrm{x}+f_\mathrm{ac}\sin\omega
  t\right)^2\bigr]^{3/2}}\;.
\end{equation}

\begin{figure}
\includegraphics[width=7cm]{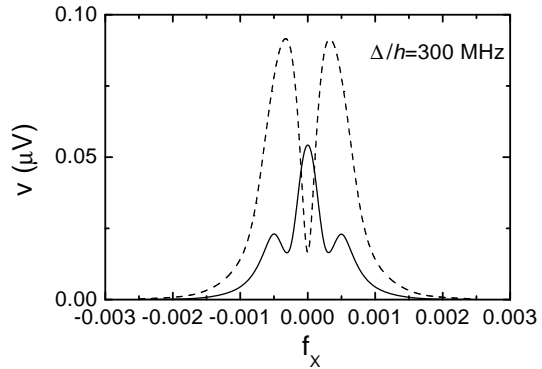}
\caption{\label{fig11} Tank voltage vs bias flux $f_\mathrm{x}$; $\Delta/h=300$~MHz,
$\Phi_\mathrm{ac}=5\cdot10^{-4}\Phi_0$. Dotted line: driving frequency
$\omega=\omega_\mathrm{T}/2$; solid line:
$\omega=\omega_\mathrm{T}/3$.}
\end{figure}

The advantage of a separate driving coil is that one can
effectively decouple the tank from the fundamental harmonic of the
bias, since the qubit signal $G(t)$ contains not only $\omega$ but
also $2\omega$, $3\omega$, etc. At $f_\mathrm{x}=0$, $G(t)$
contains only odd harmonics. This can be used to find the
degeneracy point in practice, e.g., by tuning $f_\mathrm{x}$ so
that the tank response vanishes (reaches its maximum) at frequency
$2\omega$ ($3\omega$). We have studied the higher harmonics by
solving Eq.~(\ref{V-dyn2}) numerically with
$\widetilde{\Phi}_\mathrm{ac}=0$ for $\omega=\omega_\mathrm{T}/2$
and $\omega=\omega_\mathrm{T}/3$, see Fig.~\ref{fig11}. Since the
full amplitudes contain contributions from all harmonics, at
$\omega=\omega_\mathrm{T}/2$, $f_\mathrm{x}=0$ one observes a
finite dip rather than a zero.

\section{Requirements on noise sources}
\label{noise-src}

Figures \ref{fig6} and~\ref{fig7} clearly reveal the quantum nature
of the flux qubit within a range $|\delta f_\mathrm{x}|\le
5\cdot10^{-4}$ from the degeneracy point $f_\mathrm{x}=0$.
Therefore, the unavoidable external flux noise coupled to the
qubit must be much smaller than this value. The most important
sources are the Nyquist noise
$I_\mathrm{n}=\sqrt{4k_\mathrm{B}T\!/R_\mathrm{T}}$ and the
current noise $I_\mathrm{a}$ of the preamplifier. The former
generates the qubit-flux noise $\Phi_\mathrm{n}=MI_\mathrm{n}Q\sqrt{B}$,
where $B=\omega_\mathrm{T}/2\pi Q$ is the tank bandwidth. With
$T=20$~mK and the tank parameters of Section~\ref{bit-tank}, one
gets $\Phi_\mathrm{n}\approx8\cdot10^{-6}\Phi_0$. For
$I_\mathrm{a}=10^{-14}$A$/\sqrt{\mathrm{Hz}}$, we estimate the
corresponding flux noise as
$\Phi_\mathrm{a}=MI_\mathrm{a}Q\sqrt{B}\approx7\cdot10^{-6}\Phi_0$.
Thus, the noise these sources couple to the qubit is at least two
orders smaller than the peak widths in Figs.\ \ref{fig6}
and~\ref{fig7}. On the other hand, these sources give rise to
directly detected voltage noise across the tank circuit. The
thermal tank noise is
$V_\mathrm{n}=I_\mathrm{n}\omega_\mathrm{T}L_\mathrm{T}Q\sqrt{B}\approx17.6$~nV.
The noise due to $I_\mathrm{a}$ is
$V_\mathrm{a1}=I_\mathrm{a}\omega_\mathrm{T}L_TQ\sqrt{B}\approx16$~nV.
And finally, if we take $V_\mathrm{a2}=40$~pV$/\sqrt{\mathrm{Hz}}$
for the preamplifier's own voltage noise, we get
$V_\mathrm{a2}\sqrt{B}\approx7$~nV for its contribution in the
tank bandwidth. Comparing these values with the voltage modulation
in Figs.\ \ref{fig7}, \ref{fig9}, and~\ref{fig11}, we see they are
at least several times smaller than the qubit signal.

\section{Conclusion}
We have shown that IMT can be used for low-frequency
characterization of the ground (in general: equilibrium) state of
a flux qubit. The method allows determining the tunnel splitting
between qubit states for a broad class of devices; with the term
``flux (as opposed to phase) qubit'' we stress that the two states
must differ not only in Josephson phase, but in actual magnetic
flux visible to the outside.\cite{Makhlin} The design exclusively
employs present-day technology, and the expected noise levels have
been shown not to disrupt the measurement. On the qubit time
scale, the method is a quasi-equilibrium one; as such, it can
determine $\Delta$ but not, e.g., the ``$T_2$'' dephasing
time.\cite{amb} Efforts to adapt IMT to this and related problems
such as qubit readout and control are underway.

\begin{acknowledgments}
We thank A.Yu.\ Smirnov and A.M. Zagoskin for detailed discussions.
\end{acknowledgments}

\end{document}